\documentclass[preprint,amsmath,amssymb,nofootinbib]{revtex4}

\usepackage{graphicx}% Include figure files
\usepackage{psfig}
\usepackage{dcolumn}% Align table columns on decimal point
\usepackage{bm}% bold math

\begin{document}
\title{Constraining Inverse Curvature Gravity with Supernovae}

\author{Olga Mena}
\author{Jos\'e Santiago}
\affiliation{ Fermi National Accelerator
Laboratory, P.O. Box 500,  Batavia, IL 60510, USA}
\author{Jochen Weller}
\affiliation{Department of Physics and Astronomy, University College
London, Gower Street, London WC1E 6BT, U.K. \\
Fermi National Accelerator
Laboratory, P.O. Box 500, Batavia, IL 60510, USA  }
\begin{abstract}

We show that the current accelerated expansion of the Universe
can be explained without resorting to dark energy. Models
of generalized modified gravity, with inverse powers of the 
curvature can have late time accelerating attractors without
conflicting with solar system experiments. We have solved the Friedman
equations for the full dynamical range of the evolution of the
Universe. This allows us to perform a detailed
analysis of Supernovae data in the context of such models that 
results in an excellent fit. Hence, inverse curvature gravity models
represent an example of phenomenologically viable models in
which the current acceleration of the Universe is driven by curvature
instead of dark energy. 
If we further include constraints on
the current expansion rate of the Universe from the Hubble Space Telescope and
on the age of the Universe from globular clusters, we
obtain that the matter content of the Universe is $0.07\le\omega_{\rm m}\le
0.21$ (95\% Confidence). Hence the inverse curvature gravity models
considered can
{\em not} explain the dynamics of the Universe just with a baryonic
matter component.

\vspace{0.3cm} \noindent PACS:04.50.+h, 95.36.+x \hspace{0.4cm}

\end{abstract}

\maketitle

It is now widely accepted that recent Supernovae (SNe) observations imply
that our Universe is currently experiencing a phase of accelerated
expansion~\cite{Riess:1998cb}. This seems to be independently confirmed
by observations of clusters of galaxies~\cite{Allen:04} and
the cosmic microwave background~\cite{WMAP}.
The accelerated expansion is usually explained through violations of
the strong energy condition by introducing an extra component in the
Einstein equations in the form of dark energy with an equation of
state $w<-1/3$. However, such
an explanation is plagued with theoretical and phenomenological
problems, such as the extreme fine tuning of initial conditions and the
so called coincidence problem \cite{Steinhardt} and it is therefore
natural to seek alternatives to dark energy as the source of the
acceleration. One possibility is an inhomogeneous Universe with only
local acceleration, albeit it is hard to explain natural boundary
conditions for such a local void~\cite{Kolb:2005me}. The other, that
we will elaborate on in this {\em Letter}, is modifications of
gravity that turn on only at 
very large distances \cite{Gabadadze:2004dq} or small
curvatures~\cite{Carroll:2003wy,Capozziello:2003tk}
therefore giving a geometrical origin to the accelerated
expansion of the Universe.

It was shown in~\cite{Carroll:2003wy} that a
simple modification of the gravitational action adding inverse of
curvature invariants to the Einstein-Hilbert term would naturally have
effects only at low curvatures and therefore at late cosmological
times. The simplest of such modifications includes just one single
inverse of the curvature scalar $\mu^4/R$, with $\mu$ a parameter with
dimensions of mass. This 
results in a model governed by the Einstein-Hilbert term,
\textit{i.e.} usual
gravity, for curvatures $R\gg \mu^2$ but can lead to an accelerated 
expansion at curvatures $R \lesssim \mu^2$.
This simple model is equivalent to a Brans-Dicke
theory~\cite{Carroll:2003wy}. Based on this equivalence it was  
subsequently proven by a number of authors that the model is
in conflict with solar system data~\cite{solarsystem} and is unstable
when matter is introduced~\cite{Dolgov:2003px}. This conclusion
naturally extends to generalizations of this action where the
Einstein-Hilbert 
term is supplemented with an arbitrary function of $R$, except for
particular cases that could still lead to viable
models~\cite{Dick:2003dw}. 

With this restriction in mind, the authors of~\cite{Carroll:2004de} 
discussed a
more general modification of gravity based on the following 
gravitational action
\begin{equation}
\begin{array}{lcl}
\displaystyle S &\displaystyle = &\displaystyle \frac{1}{16 \pi G}\int
    d^4x \sqrt{-g} \Big[R-\frac{\mu^{4n+2}}{(aR^2+b P + c 
    Q)^n}\Big] \\ &\displaystyle +& \displaystyle \int d^4 x \sqrt{-g}
    \mathcal{L}_M\; , 
\end{array}
\label{generalised:mod:grav}
\end{equation}
where $P\equiv R_{\mu\nu} R^{\mu\nu}$, $ Q \equiv R_{\mu\nu\rho\sigma}
R^{\mu\nu\rho\sigma}$, $G$ Newton's constant, ${\cal L}_M$ the matter
Lagrangian and $g$ the determinant of the metric.

In this generalized case the equivalence with a Brans-Dicke theory is 
not clear and a more detailed analysis of modifications of Newton's
potential has to be done to compare with solar system data. The
authors of \cite{Navarro:2005gh} computed the corrections to
Newton's law in these models as a
perturbation around Schwarzschild geometry and found that as
long as we include inverse powers of the Riemann tensor ($c\ne0$),
Newton's law is not modified in the solar system at distances shorter
than $r_c \sim 10$ pc and therefore all solar system  
experiments are well under control. Note that, as long as the Riemann
tensor is present, this result is
independent of whether we include or not inverse powers of the
scalar curvature or the Ricci tensor squared, as they vanish in the
background solution. 
This important result restricts the parameter space of 
phenomenologically relevant inverse curvature
gravity models to the ones with inverse powers of the Riemann tensor
squared present. Other constraints come from the absence of ghosts 
in the spectrum, requiring specific relations between $b$ and
$c$~\cite{Navarro:toappear}. 
Finally we restrict our analysis in this {\em Letter}
to models with $n=1$. 

Let us turn now to the cosmology of models governed by the
gravitational action (\ref{generalised:mod:grav}).
Assuming a cosmological setup
with a spatially flat Friedmann-Robertson-Walker metric,
$ds^2=-dt^2+a(t) d\vec{x}^2$, 
all models with $n=1$ can be characterized by just three parameters,
$\alpha$, $\hat{\mu}$ and $\sigma$, given in terms of the parameters
in Eq. (\ref{generalised:mod:grav}) by 
\begin{align}
\alpha\equiv & \frac{12 a+4b+4c}{12a+3b+2c}, \\
\hat{\mu} \equiv & \mu/|12a+3b+2c|^{1/6}, \\
\sigma \equiv & \mathrm{sign}(12a+3b+2c).
\end{align}
In order to write the corresponding Friedmann equation in the simplest
possible way we will use logarithmic variables, $u\equiv
\ln(H/\hat{\mu})$ and  $N \equiv \ln a$, where as usual
$H=\frac{\dot{a}}{a}$, with a dot denoting the time derivative.  
The generalized Friedmann equation in these
variables reads
\begin{equation}
u^{\prime\prime} \mathcal{P}_1(u^\prime)
+ \mathcal{P}_2(u^\prime)+ 18 \sigma \big(\mathcal{P}_3(u^\prime))^3
e^{6u}(e^{2(\bar{u}-u)}-1)=0,\label{friedmann:eq}
\end{equation}
where a prime denotes the derivative with respect to $N$ and 
we have defined the following polynomials,
\begin{align}
\mathcal{P}_1(y)&=
6 \alpha^2 y^2+24 \alpha y+32-8\alpha,
\\
\mathcal{P}_2(y)&=
15 \alpha^2 y^4+2 \alpha (50-3\alpha) y^3
\nonumber \\&
+4(40+11\alpha)y^2
+24(8-\alpha) y + 32, \\
\mathcal{P}_3(y)&=\alpha y^2+4 y+4\;.
\end{align}
The source is 
$\bar{u}
\equiv
\ln\left[ 
\bar{\omega}_r \exp(-4N) + \bar{\omega}_m \exp(-3N) \right]/2$, 
where 
we have defined the appropriately normalized
values of the energy densities \textit{today} as
\begin{equation}
\frac{8 \pi G}{3} \frac{\rho_{r,m\,0}}{\hat{\mu}^2} \equiv 
\bar{\omega}_{r,m}\;,
\end{equation} 
with $\rho_{\rm r,m}^0$ the present densities in matter and radiation and
we have exploited the fact that 
the energy-momentum tensor is still covariantly conserved.
This means that the source in
Eq. (\ref{friedmann:eq}) corresponds to the standard one
with no dark energy. 

The new Friedmann equation is no longer algebraic but a second order
non-linear differential equation. Furthermore,  
it becomes non-autonomous in the presence of sources, making
its dynamical study a formidable problem. The asymptotic behavior of
the system in vacuum was carefully  
studied in \cite{Carroll:2004de}, where it was found that, 
depending on the value of $\alpha$, but irrespective of $\sigma$,
the system has a number of  
attractors, including sometimes singularities. The same attractor and
singular points are relevant when sources are present. In that case
however, both the value of $\sigma$ and the fact that the Universe is
in a matter dominated era before the new corrections become relevant
are crucial to determine the fate of the Universe. 

A careful analysis of the dynamical behavior of the system
reveals that physically valid solutions only exist for 
certain combinations of $\alpha$ and $\sigma$. 
In order to classify the different regions, 
we define the following special values of $\alpha$: $\alpha_1 \equiv 8/9$,
$\alpha_2 \equiv 4(11-\sqrt{13})/27\approx 1.095$ and $\alpha_3 \equiv
20(2-\sqrt{3})/3 \approx 1.786$.
For $\alpha < \alpha_1$ both signs of $\sigma$ result in an acceptable
(non-singular) 
dynamical evolution, but nevertheless
in a bad fit
to Supernovae data.
For $\alpha_1 < \alpha < \alpha_2$ only $\sigma = - 1$ leads 
to an acceptable expansion history,
since for $\sigma=+1$ a singular point is violently
approached in the past. For
$\alpha_2 < \alpha$  the singular point is approached
for $\sigma = -1$, hence $\sigma = +1$ is the only physically valid
solution. In this latter case, when $\alpha_2<\alpha<\alpha_3$, 
the system goes to a stable attractor that is decelerated, thus
giving a bad fit to SNe data, for
$\alpha<32/21$ and gets accelerated for larger $\alpha$. For
$\alpha_3<\alpha$ there is no longer a stable attractor and the system
smoothly goes to a singularity in the future. That singularity occurs
earlier as $\alpha$ increases so that there is a limiting function
$\alpha_4(\bar{\omega}_m)$, 
at which the singularity is reached today. It is important to stress
that this singularity is approached in a very smooth fashion, allowing
for a phenomenologically viable behavior of the system, as opposed to
the evolution when the \textit{wrong} value of 
$\sigma$ is chosen, where the singularity is hit almost
instantaneously. Finally, for values of $\alpha\gtrsim 24.9$, there
are stable attractors again but these are never accelerated and the
resulting fit to SNe data is not acceptable. To summarize, there are
two regions that give a dynamical evolution of the system compatible
with SNe data, the {\em low} region with $\alpha_1<\alpha<\alpha_2$, for
which $\sigma=-1$, and the {\em high} region where $\alpha_2< \alpha<
\alpha_4$, for which $\sigma=+1$.

As we have emphasized it is extremely difficult to solve the dynamics 
of the system analytically. To overcome this limitation we have 
performed a comprehensive numerical study of the model resulting in
the general behavior we have outlined above.
To make things more
complicated the new Friedmann equation is extremely stiff,
due to the exponentials in the last term. 
This stiffness is directly linked to the nature of the corrections that
are negligibly small in the far past, where the curvature is much smaller
than the scale $\hat{\mu}^2$. It also makes it essentially 
impossible to numerically 
integrate it from a radiation dominated era all the way to
the present. In order to circumvent this problem, we have matched 
a perturbative
analytical solution that tracks the solution in standard Einstein
gravity in the far past to the corresponding numerical one in the
region $z \gtrsim 5$, where the analytical solution is still an extremely
good approximation, and the numerical codes can cope with the
integration. Although the matching at this point is accurate below
the 1\% level, we emphasize that it is safely above the redshift range
probed by SNe.
The approximate solution from the perturbation analysis, for $\alpha
\ne 8/9$, is given by 
\begin{equation}
H_{\mathrm{approx}}=\hat{\mu}
e^{\bar{u}}\left(1+\frac{e^{-6\bar{u}}}{36 \sigma} 
\frac{\bar{u}^{\prime\prime} \mathcal{P}_1(\bar{u}^\prime)+
\mathcal{P}_2(\bar{u}^\prime)}{\big(\mathcal{P}_3(\bar{u}^\prime)\big)^3}
\right). \label{H:approx}
\end{equation}
This is an extremely accurate solution to the full non-linear equation
as long as $z\gtrsim 5$, regardless of the values of $\alpha$ and
${\bar \omega}_m$.
At the boundaries between regions with different dynamical behavior
(including $\alpha_1$) the sensitivity to initial conditions is large
and therefore nothing conclusive can be said at these points. 
The question of sensitivity to initial
conditions is a relevant one due to the non-linear nature of Friedmann
equation. However due to the complication of any analytical study for
non-negligible sources alluded to above we will defer its study to a
future publication. In the present {\em Letter} we will contempt ourselves
with the particular solution in Eq.~(\ref{H:approx}) that we are
guaranteed tracks the standard behavior in Einstein gravity in the
past. We further explicitly confirmed, by a numerical analysis, that
our conclusions are not sensitive to the exact position of the
matching point in the past.

Once we have solved for the Hubble parameter as a function of the
scale factor, we perform a fit
to SNe data to get the allowed values of the different parameters
defining our model. In principle there is a total of five parameters defining our Universe in this framework, namely the three
parameters defining the model, $\hat{\mu}$, $\alpha$ and $\sigma$, and
the two parameters determining the sources, $\bar{\omega}_{r,m}$. 
The absolute value of the CMB temperature, however,
fixes the total radiation content of the Universe, constraining
$\bar{\omega}_r\hat{\mu}^2$. For relevant values of $\hat{\mu}$ this
constraint makes radiation irrelevant in the analysis of SNe data.
Since the intrinsic magnitude of SNe is a nuisance parameter in our
analysis, it is not possible to determine $\hat{\mu}$ as an
independent parameter with SNe only. For a standard $\Lambda$CDM Universe this
corresponds to the inability of SNe data to independently determine the
Hubble constant $H_0$. However, we
will be able to determine the value of $\hat{\mu}$ once we impose
other constraints, like the measurement of the Hubble constant by the
Hubble Key Project, $H_0=72\pm 8$ Km s$^{-1}$ 
Mpc$^{-1}$~\cite{Freedman:2000cf}. Hence, this leaves us with just three
parameters, $\alpha$, $\sigma$ and $\bar{\omega}_m$ relevant for the
analysis of SNe data and an additional nuisance parameter in terms 
of the intrinsic magnitude.

The fits are performed using the recent gold SNe data set from the last
reference in~\cite{Riess:1998cb}. The apparent magnitude is given by
$ m(z)=\mathcal{M}+5 \log \mathcal{D}_L$
where $\mathcal{M}\equiv M-5 \log \hat{\mu}+25$ and $\mathcal{D}_L
  \equiv \hat{\mu} d_L$ with $d_L \propto \int H^{-1}(z)dz$. Note
  that the parameter $\hat{\mu}$ appears in the definition of the
  magnitude compared to the usual definition involving $H_0$
  \cite{Riess:1998cb}. The important point is that $\mathcal{D}_L(z)$ can
  now be computed solely in terms of
  $\bar{\omega}_m$ and $\alpha$, 
where $\hat{\mu}$ and the intrinsic magnitude have
  been absorbed into the nuisance parameter $\mathcal{M}$ that can be
  marginalized  analytically in the probability function.

\begin{figure}[ht]
\setlength{\unitlength}{1cm}
\centerline{
\hbox{\psfig{file=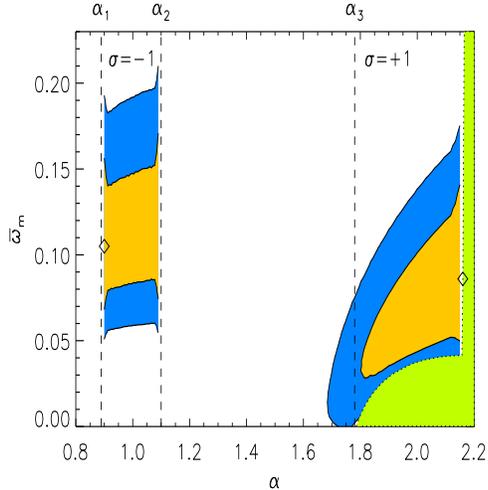,height=7cm,width=7.cm}}} 
\caption{1 and 2-$\sigma$ joint likelihoods on ${\bar\omega}_m$ and
  $\alpha$. In the {\em low} region $\sigma=-1$ whereas in the {\em
  high} region 
  $\sigma=+1$. The shaded area on the right determines the region
  $\alpha>\alpha_4$ 
  that is excluded because of a singularity being hit in the past.
The diamonds denote the maximum likelihood points.}
\label{fig:aobar}
\end{figure}

We have performed independent two parameter fits to SNe data 
for each of the {\em low} and {\em high} regions. This results in the
1- and 2-$\sigma$ 
joint likelihoods shown in Fig.\ref{fig:aobar}, 
with best fit values
given by
\begin{align}
&\mathit{low}: \, \alpha=0.9, \quad \bar{\omega}_m= 0.105, \quad
\chi^2=184.9, 
\\
&\mathit{high}: \alpha=2.15, \quad \bar{\omega}_m= 0.085, \quad
\chi^2=185.2 . 
\end{align}
For comparison 
purposes, we have also performed the fit using the standard
$\Lambda CDM$ model for a spatially flat Universe 
and absorbing $H_0$ as a nuisance parameter into ${\cal M}$, resulting in
$\chi^2=184.9$ for $156$ data points. We further show in
Fig.~\ref{fig:aobar} the points $\alpha_i$. The shaded area
on the right side, which is bordered by a dotted line, is the
exclusion zone given by $\alpha_4(\bar\omega_{\rm m})$. Note that the
contours have a sharp cut-off at $\alpha_1$, $\alpha_2$ and
$\alpha_4$. However at $\alpha_3$ there is no singularity hit
violently and the $2-\sigma$ contour of the {\em high} region extends
below $\alpha_3$. In the {\em low} region we obtain $\bar\omega_{\rm m}
= 0.122 \pm 0.034$ after marginalization over $\alpha$ and in the {\em
  high} region $\bar\omega_{\rm m} = 0.075 \pm 0.031$.
Note that our best fit points in both regions are close to the borders
of the allowed region. This is because within the regions there is a
smooth behavior of the likelihood, and only the dynamics of the
system cuts off the likelihood space if certain parameter values are
reached.

If we
additionally apply the
HST measurement of $H_0$, \cite{Freedman:2000cf} we can determine
$\hat{\mu}$ and the matter 
content $\omega_{\rm
  m} = \Omega_{\rm m}h^2$, with $H_0 = 100 h\,{\rm km/s/Mpc}$.
Finally we can restrict the allowed region in
$\omega_{\rm m}-\hat{\mu}$ a little bit more by imposing a prior on
the age of the Universe with a mean of $t_0 = 13.4$ Gyrs and a 95\%
confidence lower limit 
of $11.2$ Gyrs~\cite{Krauss:03} . 
\begin{figure*}[ht]
\setlength{\unitlength}{1cm}
\centerline{\hbox{\psfig{file=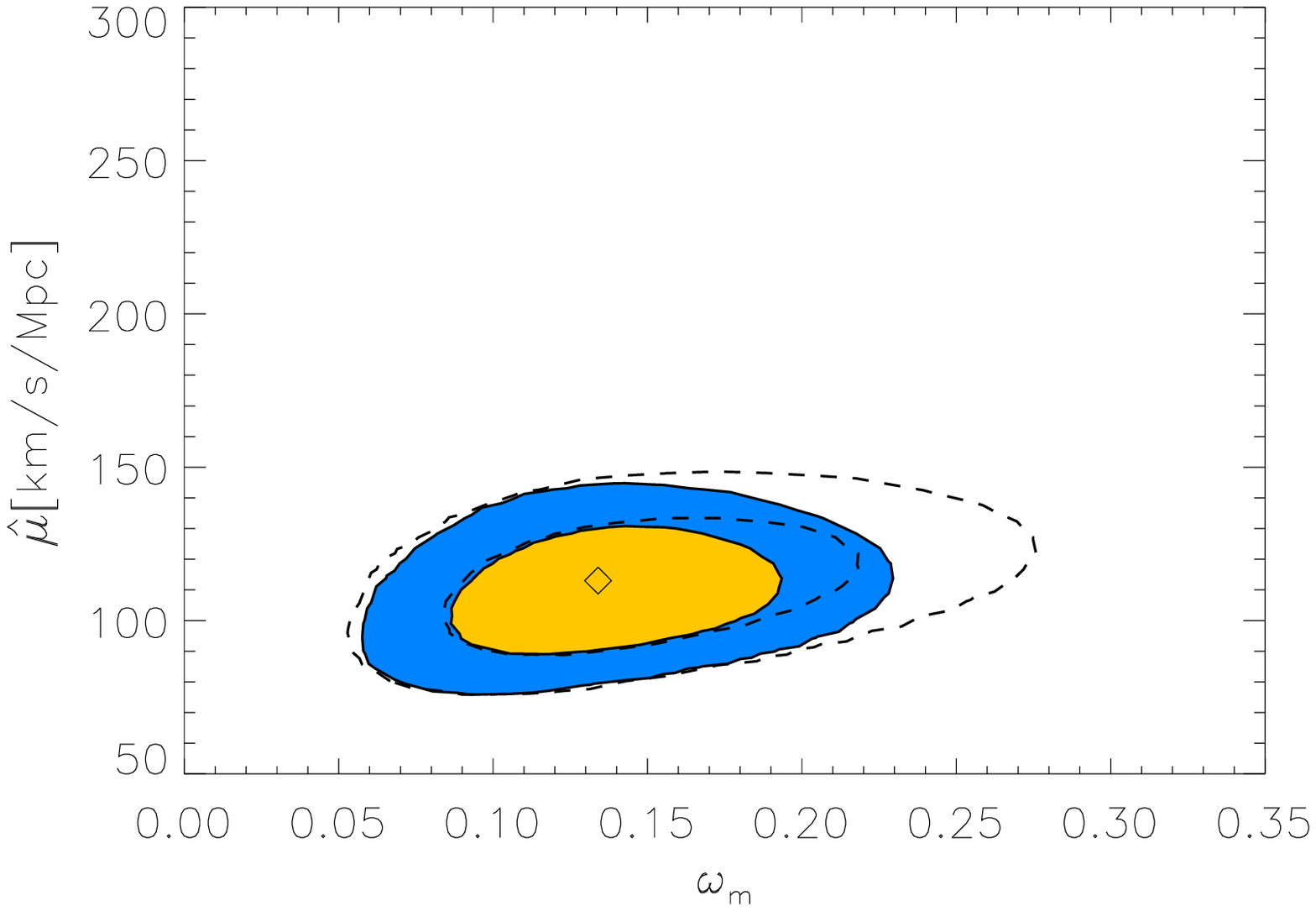,height=7cm,width=7.cm}
\psfig{file=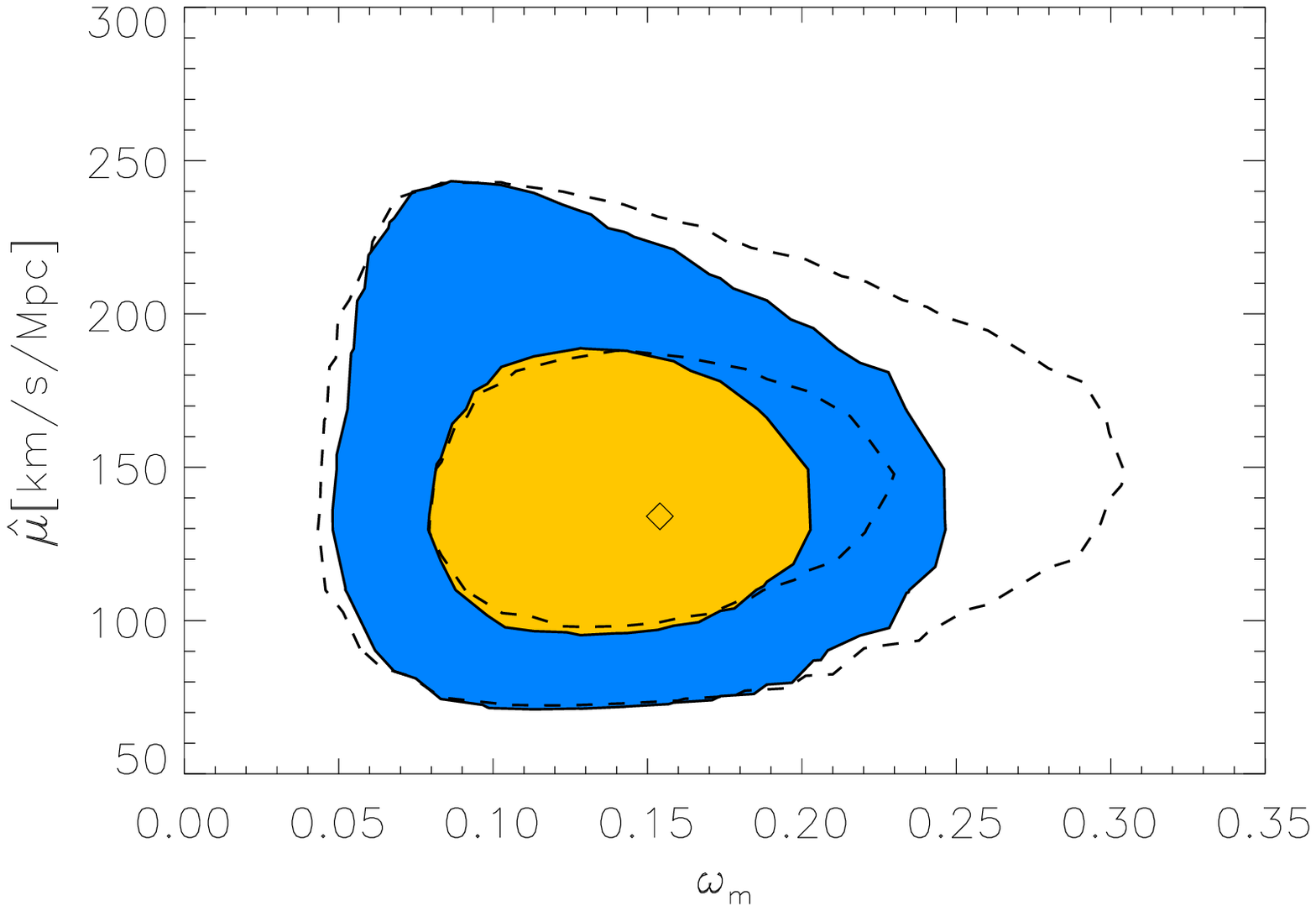,height=7cm,width=7.cm}}}
\caption{The 1- and 2-$\sigma$ joint likelihoods in the $\omega_{\rm
    m}-\hat{\mu}$ plane, when additional priors on $H_0$ and the age
    of the Universe are
    imposed. On the left for the {\em low} region and on 
  the right for the {\em high} region. Diamonds are the maximum
    likelihoods. Further, the dashed contours are the joint
    likelihoods if we impose only the $H_0$ prior.}  
\label{fig:ommu}
\end{figure*}
In Fig.~\ref{fig:ommu} we show the joint 1- and 2-$\sigma$ likelihoods
in the $\omega_m$-$\hat{\mu}$ plane, with both priors imposed
(solid line) and without imposing the age of the Universe prior
(dashed line). On the left for the {\em low}
region and on the right for the \textit{high} region. First we recognize
that $\hat{\mu}$ is roughly twice the size of the Hubble constant
$H_0$. If we further marginalize over $\hat{\mu}$ the physical matter
content in the Universe is $\omega_{\rm m}=0.14 \pm 0.03$ and
$\omega_{\rm m}=0.14 \pm 0.04$ in the \textit{low} and {\em high}
regions, respectively.
Note that the matter content in the 
budget of the Universe is clearly higher than the measured baryonic
content. Overall we find $0.07 \leq \omega_{\rm m}\leq 0.21$ at the 95\%
confidence level. If we compare this number with the results from Big
Bang Nucleosynthesis $\omega_{\rm b} = 0.0214 \pm 0.0020$
\cite{Kirkman:03} it is clear that we require a dark matter component
to explain the data.

Other cosmological probes such as clusters of galaxies and CMB could
further constrain these models. However such an analysis is beyond the
scope of this {\em Letter} since it requires a detailed re-calculation
of, e.g. cluster potentials and CMB perturbations for the models
discussed here.

{\em Summary}: We have studied the viability of a geometrical explanation
for the present acceleration of the Universe. This is possible if the
Einstein-Hilbert action is supplemented with new terms that are
negligibly small at high cosmological curvatures but become relevant
when the curvature of the Universe gets smaller. Despite the
phenomenological problems of the simplest models, it has been shown
that there exists a broad class of modifications of gravity that are
phenomenologically viable and have accelerated attractors at
late-times. In this {\em Letter} we have performed a detailed numerical 
analysis of the dynamics of these models. We emphasize that this hard
numerical problem has not been solved previously. The result of this
analysis allowed us to compare inverse curvature gravity with
Supernovae data. We found that SNe data can be fitted
in our model without the need of any dark energy and getting
meaningful constraints in the free parameters. We further have shown
that these models still require a dark matter component. Of course
this latter conclusion does not need to hold for more general models, for
instance those with $n\neq 1$. We are planning to study more general
models and their implications for dark matter in the near
future. However, we would like to emphasize the generality of our
study. We have parameterized \textit{all} models governed by
Eq. (\ref{generalised:mod:grav}) with $n=1$.
Finally we are currently extending this analysis to CMB and cluster
datasets, a non-trivial task. This will further constrain these models
and maybe even distinguish them from dark energy.

{\em Acknowledgments}: It is a pleasure to thank R. Battye,
G. Bertone, S. Dodelson, A.~Lewis, M. Liguori, I.Navarro for useful
conversations.  
This work is supported by DOE and NASA grant NAG 5-10842.

\end{document}